\documentclass[12pt]{article} 
\usepackage[latin1]{inputenc}
 \usepackage{graphicx}
\usepackage{amssymb}
\usepackage{amsmath}

\newcommand{\longvec}{\stackrel{\longrightarrow}}

\begin{document}
\title{Arrangement of a 4Pi microscope for reducing the confocal detection volume with two-photon excitation}

\author{Nicolas Sandeau\footnote{E-mail: nicolas.sandeau@fresnel.fr} and Hugues Giovannini\footnote{E-mail: hugues.giovannini@fresnel.fr; Tel: +33 491 28 80 66; Fax: +33 491 28 80 67}}
\date{\small{Received 16 November 2005; received in revised form 7 February 2006; accepted 8 February 2006}}
\maketitle
\begin{center}
\emph{Institut Fresnel, UMR 6133 CNRS, Université Paul Cézanne Aix-Marseille III, F-13397 Marseille cedex 20, France
}
\end{center}

\noindent The main advantage of two-photon fluorescence confocal microscopy is the low absorption obtained with live tissues at the wavelengths of operation. However, the resolution of two-photon fluorescence confocal microscopes is lower than in the case of one-photon excitation. The 4Pi microscope type C working in two-photon regime, in which the excitation beams are coherently superimposed and, simultaneously, the emitted beams are also coherently added, has shown to be a good solution for increasing the resolution along the optical axis and for reducing the amplitude of the side lobes of the point spread function. However, the resolution in the transverse plane is poorer than in the case of one-photon excitation due to the larger wavelength involved in the two-photon fluorescence process. In this paper we show that a particular arrangement of the 4Pi microscope, referenced as 4Pi' microscope, is a solution for obtaining a lateral resolution in the two-photon regime similar or even better to that obtained with 4Pi microscopes working in the one-photon excitation regime.\\


\section{Introduction}
Strong efforts have been made in the last decade to improve the resolution of fluorescence microscopes. Indeed, localizing marked species with sub-wavelength accuracy gives precious information for cell biology applications. In particular, in fluorescence correlation spectroscopy experiments made with confocal microscopes, varying the detection volume is the key task for studying, at different scales, molecular mechanisms inside cells \cite{Wawrezinieck2004,Masuda2005,Wawrezinieck2005}. For reducing the lateral extent of the detection volume, high numerical aperture immersion objectives have been developed \cite{Martini2002}. However, the axial extent of the point spread function (PSF) of conventional confocal microscopes remains about four times larger than its lateral extent. To solve this problem, various solutions, mostly based on the use of interference phenomena, have been proposed \cite{Sheppard1991,Stelzer1994,Nagorni2001,Gustafsson2002,Lenne2002}. In particular the coherent superposition of the excitation wavefronts and that of the emission wavefronts passing through two opposing lenses, has led to the development of the 4Pi microscope \cite{Hell1992c,Hell1992}. It has been shown that, with 4Pi microscopes working with one-photon excitation, the axial resolution can be improved by a factor 3 to 7 over that of confocal microscopes and related systems \cite{Egner2005}. However, with this technique, the focal maximum is also accompanied by interference side lobes whose maximum intensity exceeds $50\%$ of the maximum intensity in the focal point. In this case classical image deconvolution algorithms do not work properly and the resolution along the optical axis is not improved. To overcome this difficulty, various solutions have been proposed. Among them one can cite the 4Pi type C microscope with two-photon excitation \cite{Denk1990,Sheppard1990,Hell1992a} . In this set-up, two opposite microscope objectives are used to illuminate coherently the fluorescent sample from both sides and, simultaneously, to add coherently the two emitted beams. The PSF is the result of the superposition of two systems of fringes: the one produced by the pump beams, the other produced by the emitted beams. The strong difference between the pump wavelength and the wavelength of luminescence in the two-photon excitation regime leads to different intensity spatial distributions of the fringes along the optical axis. The consequence is that a strong reduction of the amplitude of the side lobes of the PSF is obtained. This is a very interesting solution which strongly improves the resolution along the optical axis while preserving the main advantage of two-photon excitation which is the low absorption of live tissues at the wavelength of operation \cite{Denk1990,Gugel2004}. However, due to the larger wavelength of the pump beams used in the two-photon excitation regime, the transverse resolution is worse than in the case of one-photon excitation \cite{Sheppard1990,Hell1992a}. Other solutions based on the used of variable density filters for shaping the axial component of the PSF have also been proposed \cite{Martinez-Corral2003}.\\
\indent\hspace{0.5cm}Recently it has been shown theoretically that a particular arrangement of 4Pi microscope, referenced as 4Pi' microscope \cite{Sandeau2005a}, made possible an increase of the lateral resolution in one-photon excitation regime. In the present paper we extend the domain of application of the 4Pi' microscope. We describe the vector model that can be used to compute the image of a dipole through the 4Pi' microscope. Thanks to numerical simulations based on this model, we show that the 4Pi' microscope working in the two-photon regime cumulates the advantages of the 4Pi microscope working with one-photon regime and those given by two-photon excitation. In particular numerical, calculations show that the excitation volume of the 4Pi' type C microscope working with two-photon excitation is comparable to or even smaller than the excitation volume obtained with  4Pi type C microscopes working in the one-photon excitation regime. The main advantage of this solution is that it keeps a high resolution when the pinhole size increases, leading to high signal-to-noise ratio for practical applications.

\section{Set-up}

\begin{figure}[!b]
\begin{center}	\includegraphics[width=10 cm]{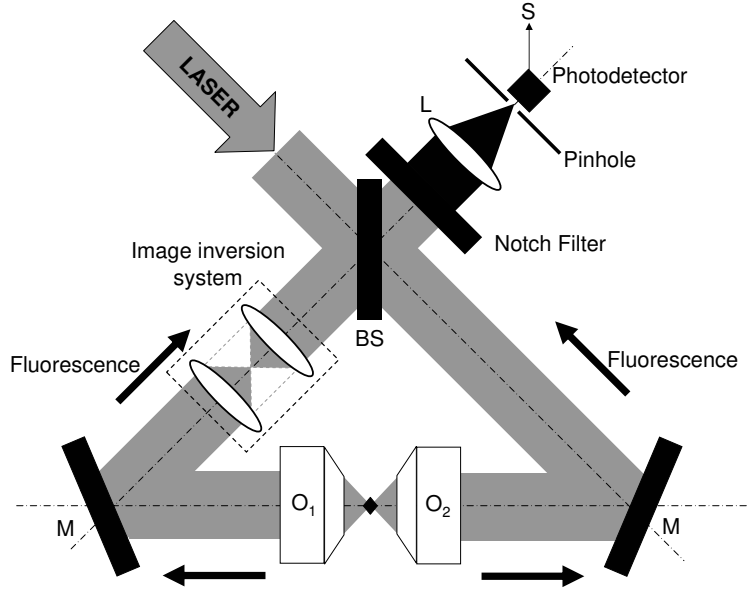}\caption{\emph{Set up of the 4Pi' microscope. With respect to the classical 4Pi microscope, an image inversion system is added in one arm of the interferometer. The optical path difference of the interferometer is assumed to be equal to 0 for the incident wavelength and for the emission wavelength. M are mirrors. BS is a beam splitter.}}\label{4pimtriangle}
\end{center}
\end{figure}
In 4Pi microscopes the axial extent of the PSF is reduced, with respect to classical confocal microscopes, by taking advantage of the variation of the optical path difference (OPD) along the optical axis between the pump beams (case of 4Pi type A microscopes), between the emitted beams (case of 4Pi type B microscopes) or between both the pump beams and the emitted beams (case of 4Pi type C microscopes) \cite{Hell1992}.The displacement of the luminescent source in the transverse direction has no influence on the OPD between the emitted beams. Thus, for identical microscope objectives and similar exit pupil diameters, the resolution is improved only along the optical axis and the lateral extent of the PSF of 4Pi microscopes is identical to that of classical confocal microscopes working at the same excitation and emission wavelengths. Recently a configuration of the 4Pi microscope, called 4Pi' microscope, has been proposed \cite{Sandeau2005a}. In this microscope an interference phenomenon is produced by the displacement of the luminescent source in the transverse direction. This interference phenomenon is used to reduce the lateral extent of the PSF. The set-up is represented in Fig. \ref{4pimtriangle}. It is based on a 4Pi microscope in which an image inversion has been added in one arm.
\begin{figure}[!tb]
\begin{center}	\includegraphics[width=9 cm]{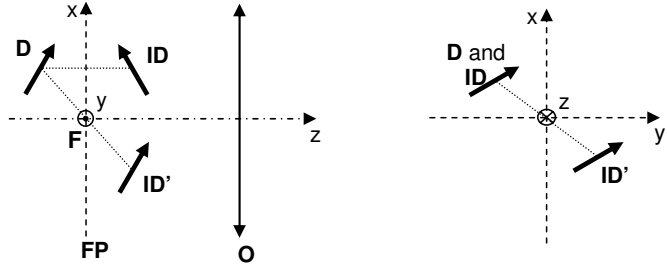}\caption{\emph{Equivalent optical schemes, for emission, of the 4Pi microscope and of the 4Pi' microscope. ID is the virtual source in the 4Pi microscope. ID' is the virtual source in the 4Pi' microscope. z is the optical axis of the system, O is the microscope objective. The two microscope objectives O1 and O2 of the set-up of Fig. \ref{4pimtriangle} are assumed to be identical. (a) View in a plane containing the optical axis. (b) View in a plane orthogonal to the optical axis. F is the common focus of O. FP is the focal plane of O.}}\label{dipoles}
\end{center}
\end{figure}
The pump laser is assumed to be transverse monomode TEM$_{00}$ in order to ensure the spatial coherence of the two incident beams which interfere as in a classical 4Pi microscope. The beam is expanded in order to give a constant illumination within the aperture of the objectives. We assume that the OPD between the two incident beams is equal to 0 in the common focal focus F of the microscope objectives. We also assume that the chromatic dispersion in the two arms is perfectly compensated. In order to point out the differences between the 4Pi microscope and the 4Pi' microscope we have represented, in Fig. \ref{dipoles}, the equivalent optical schemes of the two arrangements. We have considered a single dipole source placed in the vicinity of the common focus F of the two microscope objectives. In the 4Pi microscope and in the 4Pi' microscope the two beams emitted by dipole D pass through the objectives. In both microscopes the displacement of dipole D from focus F creates a second virtual source. However, the location of this virtual source depends on the arrangement (see Fig. \ref{dipoles}). In the 4Pi microscope the two beams are emitted by dipole D and by the virtual source ID. In the 4Pi' microscope the two beams are emitted by dipole D and by the virtual source ID'. D and ID are symmetric with respect to focal the plane FP of the microscope objectives, while D and ID' are symmetric with respect to focus F. All the differences between the 4Pi microscope and the 4Pi' microscope lie on these two different symmetries.

\section{Vector model}

\begin{figure}[!ht]
\begin{center}	\includegraphics[width=\textwidth]{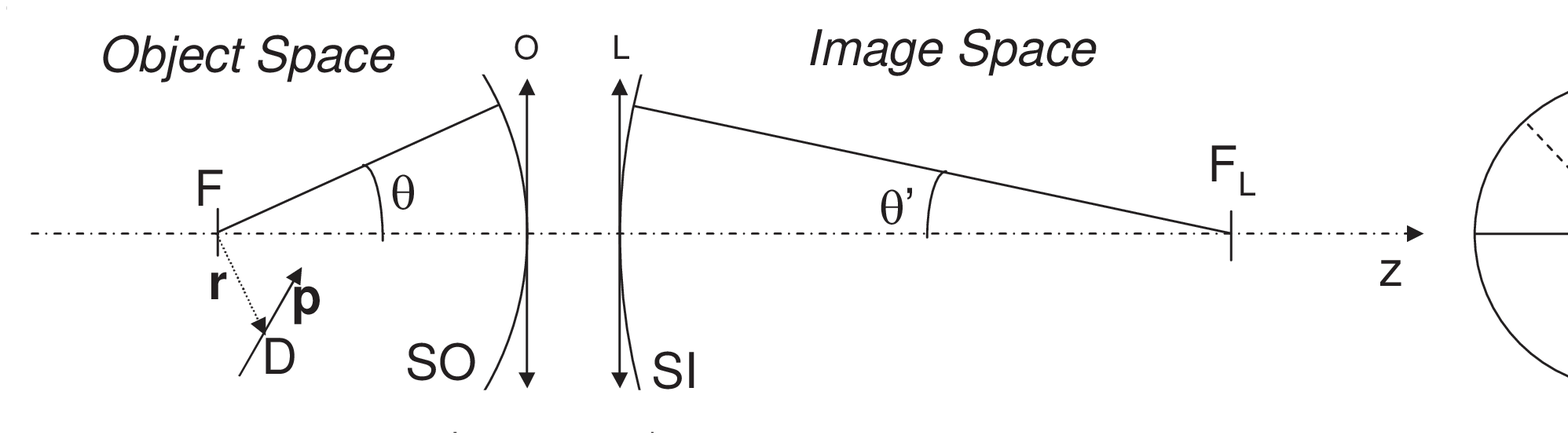}\caption{\emph{Schematic of the equivalent optical scheme of the set-up of Figure 1. O is a microscope objective identical to O1 and O2. For seek of simplicity, the distance between L and O is assumed to be equal to zero.}}\label{micro}
\end{center}
\end{figure}

For comparing the CEF of the 4Pi microscope with that of the 4Pi' microscope, for a source located around focus F, taking into account the vector nature of the dipolar emission, it is necessary to use a complete description of the image formation through the microscope objectives. We have calculated the collection efficiency function CEF(\textbf{p},\textbf{r}) for different orientations and different positions of dipole D, where \textbf{p} is the dipolar moment of D and $\mathbf{r}=\longvec{\mathtt{FD}}$. CEF(\textbf{p},\textbf{r}) is the sum of the intensities in each point of detector's surface S:
\begin{equation}\label{equ1}
CEF\left( {{\bf p},{\bf r}} \right) =\iint\limits_{S}{\left\| {{\bf E}\left( {{\bf R},{\bf p},{\bf r}} \right) + {\bf E}\left( {{\bf R},{\bf \hat p}{\bf ,\hat r}} \right)} \right\|^2 dS} \ ,
\end{equation}
with
\begin{equation}\label{equ2}
\left\| {{\bf E_1} + {\bf E_2}} \right\|^2  = \left( {{\bf E_1} + {\bf E_2}} \right) \cdot \left( {{\bf E_1}^*  + {\bf E_2}^* } \right)\ ,
\end{equation}
where \textbf{E}(\textbf{R},\textbf{p},\textbf{r}) is the electric field emitted by dipole D and \textbf{E}(\textbf{R},$\bf\hat p$,$\bf\hat r$) is the electric field emitted by ID (case of the 4Pi microscope) or ID' (case of the 4Pi' microscope) at point M of detector's surface defined by the vector $\mathbf{R}=\longvec{\mathtt{F_LM}}$ (see Fig. \ref{micro}). In eq. (\ref{equ2}), \textbf{E*} designs the complex conjugate of vector \textbf{E}. For a classical 4Pi microscope $\bf\hat p=\left(p_x,p_y,-p_z\right)$ and $\bf\hat r=\left(r_x,r_y,-r_z\right)$ whereas for a 4Pi' microscope $\bf\hat p=\bf p$ and $\bf\hat r=-\bf r$. For calculating \textbf{E}(\textbf{p},\textbf{r}) and \textbf{E}($\bf\hat p$,$\bf\hat r$) we use a vector method described by Richards et al. \cite{Richards1959} and later applied by Enderlein et al. \cite{Enderlein2000,Bohmer2003}. The electric field $\mathbf{E_0}$ emitted by dipole D is calculated on a reference sphere SO centered on F (see Fig. \ref{micro}):
\begin{equation}\label{equ3}
E_0 \left( {{\bf p},{\bf r}} \right) \propto \frac{{\left( {{\bf p}_ \bot   \cdot {\bf e}_{{\bf rd}} } \right)}}{{{\rm r}_{\rm d} }}\ \exp\left[i\:knr_d\right]{\bf e}_{{\bf \theta d}}\ ,
\end{equation}
with
\begin{equation}\label{equ4}
{\bf e}_{{\bf \theta d}}  = \left( {{\bf p} \cdot {\bf e}_{{\bf rd}} } \right){\bf p}_ \bot   - \left( {{\bf p}_ \bot   \cdot {\bf e}_{{\bf rd}} } \right){\bf p}\ ,
\end{equation}
\begin{equation}\label{equ5}
{\bf p}_ \bot   = \frac{{{\bf e}_{{\bf rd}}  - \left( {{\bf p} \cdot {\bf e}_{{\bf rd}} } \right){\bf p}}}{{\sqrt {1 - \left( {{\bf p} \cdot {\bf e}_{{\bf rd}} } \right)^2 } }}\ ,
\end{equation}
\begin{equation}\label{equ6}
{\bf e}_{{\bf rd}}  = \frac{{f{\bf e}_{\bf r}  - {\bf r}}}{{r_d }}\ ,
\end{equation}
\begin{equation}\label{equ7}
r_d  = \sqrt {\left( {f^2  - 2f\left( {{\bf e}_{\bf r}  \cdot {\bf r}} \right) + r^2 } \right)}\ ,
\end{equation}
where \emph{f} is the focal length of the microscope objective. The system of two lenses transforms the reference sphere SO in the object space, in a reference sphere SI in the image space, centred on focus $F_L$ of lens L. In this transformation $\theta$ is changed into $-\theta'$ and the spherical base $\left(\bf{e_r},\bf{e_\theta},\bf{e_\varphi}\right)$ is changed into $\left(\bf{e'_r},\bf{e'_\theta},\bf{e_\varphi}\right)$, where $\theta$ and $\theta'$ are the polar angles in the object space and in the image space, respectively, and is the azimuth angle (see Fig. \ref{micro}). We have:
\begin{equation}\label{equ8}
\theta ' = arcsin\left( {\frac{{n\sin \theta }}{m}} \right)\ ,
\end{equation}
where m is the magnification of the microscope and n is the refractive index in the object space. The refractive index in the image space is assumed to be equal to 1.
\begin{equation}\label{equ9}
{\bf e}_{\bf r}  = \left( {\begin{array}{*{20}c}
   {\sin \left( \theta  \right)\cos \left( \varphi  \right)}  \\
   {\sin \left( \theta  \right)\sin \left( \varphi  \right)}  \\
   {\cos \left( \theta  \right)}  \\
\end{array}} \right),\ {\bf e}_{\bf \theta }  = \left( {\begin{array}{*{20}c}
   {\cos \left( \theta  \right)\cos \left( \varphi  \right)}  \\
   {\cos \left( \theta  \right)\sin \left( \varphi  \right)}  \\
   { - \sin \left( \theta  \right)}  \\
\end{array}} \right),\ {\bf e}_\varphi   = \left( {\begin{array}{*{20}c}
   { - sin\left( \varphi  \right)}  \\
   {cos\left( \varphi  \right)}  \\
   0  \\
\end{array}} \right)
\end{equation}
\begin{equation}\label{equ10}
{\bf e'}_{\bf r}  = \left( {\begin{array}{*{20}c}
   {-\sin \left( \theta'  \right)\cos \left( \varphi  \right)}  \\
   {-\sin \left( \theta'  \right)\sin \left( \varphi  \right)}  \\
   {\cos \left( \theta'  \right)}  \\
\end{array}} \right),\ {\bf e'}_{\bf \theta }  = \left( {\begin{array}{*{20}c}
   {\cos \left( \theta'  \right)\cos \left( \varphi  \right)}  \\
   {\cos \left( \theta'  \right)\sin \left( \varphi  \right)}  \\
   {\sin \left( \theta'  \right)}  \\
\end{array}} \right)%,\ {\bf e}_\varphi   = \left( {\begin{array}{*{20}c}
\end{equation}
Thus, the electric field on SI is given by:
\begin{equation}\label{equ11}
{\bf E}_{\bf i} \left( {{\bf p},{\bf r}} \right) = G\left\{ {\left( {{\bf E}_{\bf o}  \cdot {\bf e}_{\bf r} } \right){\bf e}_{\bf r}^{\bf '}  + \left( {{\bf E}_{\bf o}  \cdot {\bf e}_\theta  } \right){\bf e}_\theta ^{\bf '}  + \left( {{\bf E}_{\bf o}  \cdot {\bf e}_\varphi  } \right){\bf e}_\varphi  } \right\}\ ,
\end{equation}
where G is the corrected magnification introduced by Wolf et al. \cite{Richards1959} to be accordance with the geometrical optics intensity law, with:
\begin{equation}\label{equ12}
G = m\sqrt {\frac{{\cos \theta '}}{{n\cos \theta }}}\ .
\end{equation}
The electric field in the image space near the focal plane of lens L is given by:
\begin{equation}\label{equ13}
{\bf E}\left( {{\bf R},{\bf p},{\bf r}} \right) =\iint\limits_{\Omega}{{{\bf E}_{\bf i} \left( {{\bf p},{\bf r}} \right)\exp \left[ {i\left( {{\bf k} \cdot {\bf R}} \right)} \right]d{\bf k}}}\ ,
\end{equation}
where $\Omega$ is the solid angle delimited by the numerical aperture of the objective and \textbf{k} the wave vector given by:
\begin{equation}\label{equ14}
{\bf k} = \frac{{2\pi }}{\lambda }\frac{{f{\bf e}_{\bf r}^{\bf '}  - {\bf r}}}{{\sqrt {f^2  - 2f\left( {{\bf e}_{\bf r}^{\bf '}  \cdot {\bf r}} \right) + r^2 } }}
\end{equation}
The CEF calculated using Eq. (\ref{equ1}) is averaged, for each position of the dipole emitter, over all possible dipole orientations, corresponding to imaging fluorescing molecules with a rotation diffusion much faster than the fluorescence lifetime. Thus, we can calculate ACEF(\textbf{r}) which is the averaged CEF with:
\begin{equation}\label{equ15}
ACEF\left( {\bf r} \right) = \mathop{{\int\!\!\!\!\!\int}\mkern-16mu O} 
 {CEF\left( {{\bf p},{\bf r}} \right)d{\bf p}}\ .
\end{equation}
In fluorescence microscopy experiments the samples are illuminated by the pump beam. In this case the resolution is determined by the molecule-detection efficiency function (MDEF) with:
\begin{equation}\label{equ16}
MDEF(\mathbf{r})=ACEF(\mathbf{r}).EEF(\mathbf{r})\ ,
\end{equation}
where EEF(\textbf{r}) is the Emission Efficiency Function at \textbf{r}. We have:
\begin{equation}\label{equ17}
EEF(\mathbf{r})=I_e(\mathbf{r})
\end{equation}
for one-photon excitation and
\begin{equation}\label{equ18}
EEF(\mathbf{r})=I_e^2(\mathbf{r})
\end{equation}
for two-photon excitation, where $I_e(\mathbf{r})$ is the intensity of the pump beam at \textbf{r}.

\section{Numerical simulations}

In order to study a case of practical interest, we have considered the case of Oregon Green (Molecular Probes, Eugene, OR) fluorophores that can be pumped at 976 nm (=2x488 nm) in the two-photon excitation regime and at 488 nm in the one-photon excitation regime.
\begin{figure}[!hb]
\begin{center}	\includegraphics[width=10 cm]{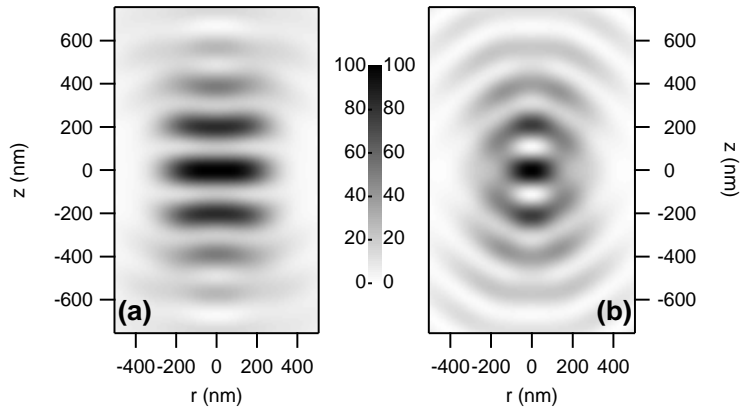}\caption{\emph{Averaged collection efficiency function ACEF. Normalized results. (a) 4Pi microscope. (b) 4Pi' microscope.}}\label{ACEF}
\end{center}
\end{figure}
The wavelength of emission of these fluorophores is around 525 nm. These excitations can be produced, at 488 nm, by an Argon ion laser and at 976 nm by a pulsed laser diode (for example 980 LDH-P, PicoQuant, Berlin, GER). We have studied the cases of one-photon excitation and two-photon excitation for the 4Pi microscope and for the 4Pi' microscope. In order to take into account the broad fluorescence spectrum whose width is around 30 nm, we have made a numerical calculation by summing incoherently ACEF(\textbf{r}) given by Eq. (\ref{equ15}) for all wavelengths of the interval [510 nm, 540 nm]. The results obtained have shown to be very similar to the results obtained in the monochromatic case at the mean wavelength  $\lambda$=525 nm. For this reason and for reducing the calculation time, we have considered a monochromatic emission. For the calculations we have considered an oil (n=1.52) immersion microscope objective with an effective numerical aperture NA=1.3, a magnification m=40 and a pinhole diameter equal to 20  $\mu$m. In this case the size of the pinhole equals the size of the Airy disk. Using Eq. (\ref{equ15}) we have computed the longitudinal section of ACEF(\textbf{r}) for the 4Pi and for the 4Pi' microscope. The results are represented in Fig. \ref{ACEF}. One can see the differences between the distributions of the ACEF due to the different symmetries introduced by the two arrangements and described in Fig. \ref{dipoles}.
\begin{figure}[!hb]
\begin{center}	\includegraphics[width=10 cm]{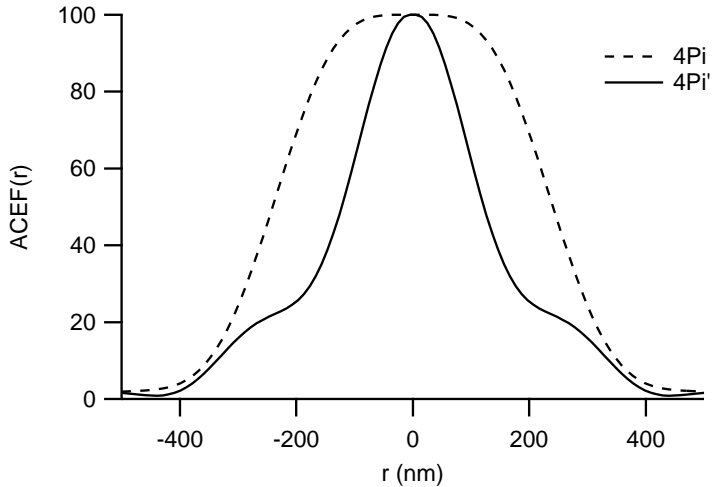}\caption{\emph{Section of ACEF(r) in the common focal plane of the microscope objectives. Normalised results.}}\label{ACEFcoup}
\end{center}
\end{figure}\\
\indent\hspace{0.5cm}Fig. \ref{ACEFcoup} shows the section of ACEF(\textbf{r}) in the focal plane. One can see that a significant improvement of the resolution is obtained with the 4Pi' microscope. The strength of this effect depends on the pinhole size and vanishes completely for very small pinhole sizes. No new spatial frequencies are introduced by the method. This effect is due to the symmetry of the interferometer and to the detection through a pinhole. A more detailed analysis of this effect can be found in Ref. \cite{Sandeau2005a}. The more the pinhole diameter increases, the more the improvement obtained with the 4Pi' microscope increases. The curves of Fig. \ref{ACEF} and of Fig. \ref{ACEFcoup} are representative of the case of a pinhole diameter equal to the diameter of the Airy disk. With respect to solutions based on the use of variable density filters in the pinhole, or based on the use of pinholes much smaller than the size of the Airy disk, the solution presented in this paper offers a main advantage : it keeps a high lateral resolution even when the pinhole size increases, leading to a better signal-to-noise ratio in practical applications. In order to calculate the emission efficiency function in the one-photon and in the two-photon excitation regimes, we have considered an incident constant illumination at 488 nm and at 976 nm on the surfaces of the microscope objectives. The beams are polarized along x axis. 
\begin{figure}[!t]
\begin{center}	\includegraphics[width=10 cm]{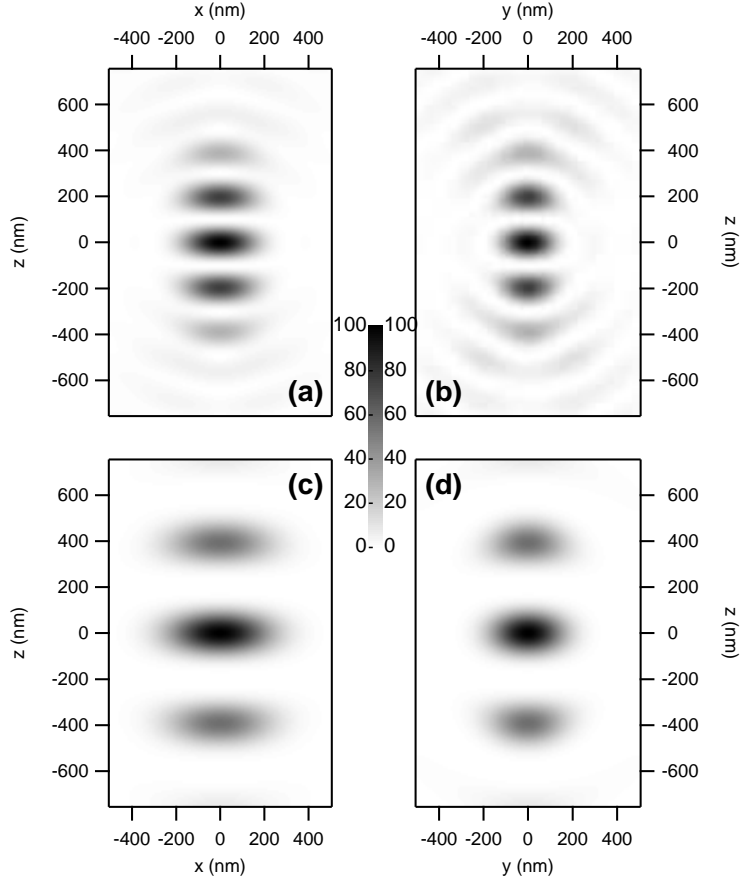}\caption{\emph{Longitudinal sections the EEF. Normalised results. The beams are polarized along x axis. (a) and (b): excitation wavelength 488 nm. (c) and (d): excitation wavelength 976 nm.}}\label{EEF}
\end{center}
\end{figure}
Fig. \ref{EEF} represents the section of EEF(\textbf{r}) given by Eq. (\ref{equ17}) and Eq. (\ref{equ18}) in two perpendicular longitudinal planes. In order to quantify the improvement obtained with a 4Pi' microscope working in the two-photon regime in terms of resolution along both transverse and longitudinal directions, we have computed the Molecule Detection Efficiency Function.
\begin{figure}[!p]
\begin{center}	\includegraphics[width=9.2 cm]{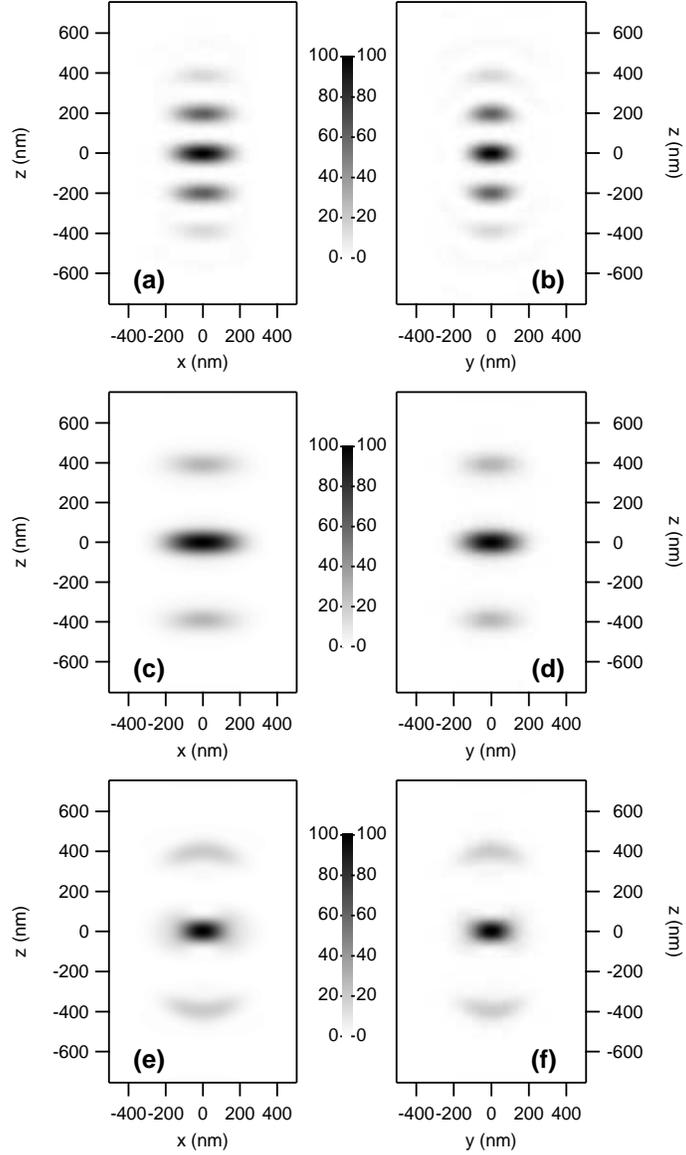}\caption{\emph{Longitudinal sections of the MDEF. Normalised results. The incident beam is polarized along x axis. Fluorescence wavelength is $\lambda=$525 nm. (a) and (b): case of a 4Pi microscope with one-photon excitation at 488 nm. (c) and (d): case of a 4Pi microscope with two-photon excitation at 976 nm. (e) and (f): case of a 4Pi' microscope with two-photon excitation at 976 nm.}}\label{MDEF}
\end{center}
\end{figure}
In Fig. \ref{MDEF} we have calculated the MDEF in different cases. For the calculations we have considered the approach of pairing coherent illumination with coherent detection, referred to as "4Pi microscopy of type C", which has proved to be a powerful solution for reducing the amplitude of the side lobes of the MDEF \cite{Gugel2004}. MDEF(\textbf{r}) was calculated using Eq. (\ref{equ16}). One can see that the lateral extent of the MDEF of a 4Pi' microscope working in the two-photon excitation regime is almost two times smaller than the lateral extent of the MDEF of a 4Pi microscope working in the same conditions.\\
\begin{figure}[!ht]
\begin{center}	\includegraphics[width=10 cm]{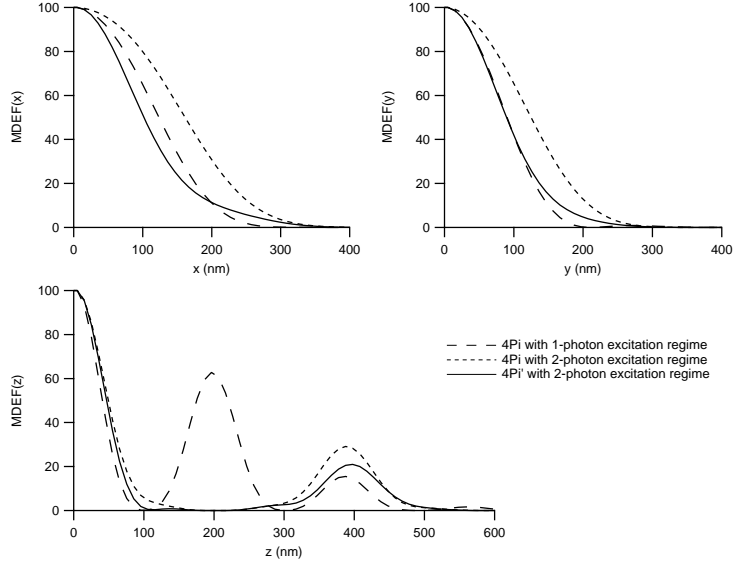}\caption{\emph{Sections of the MDEF along x, y and z axis. Normalized results.}}\label{MDEFcoup}
\end{center}
\end{figure}

\indent\hspace{0.5cm}From the curves of Fig. \ref{MDEFcoup}, one can see that, for a 4Pi' microscope working in the two-photon excitation regime, the amplitude of the side-lobes of the MDE along the optical axis is even smaller than the one obtained with the classical 4Pi microscope working in the same conditions. But the main point is that the lateral resolution obtained with the 4Pi' type C microscope working in the two-photon regime is similar or even better to the one obtained with the 4Pi type C microscope working with one-photon excitation. Similar calculations made with different numerical apertures and different wavelengths lead to the same conclusions. Moreover, one can notice that, for a pump beam linearly polarized, the section of the detection volume in the focal plane is more symmetric with the 4Pi' microscope than with the 4Pi microscope.

\section{Conclusion}

The numerical results presented in this paper have shown that the lateral resolution obtained with the 4Pi' type C microscope working in the two-photon excitation regime is comparable to that obtained with classical confocal microscopes working in the one-photon excitation regime. Moreover the amplitude of the side lobes obtained with the 4Pi' type C microscope with two-photon excitation is comparable to or even smaller than the amplitude of the side lobes obtained with the 4Pi type C microscope working in the same conditions. The 4Pi' microscope, which is suitable for scanning setups, offers the advantages given by the use of two-photon excitation, but with a transverse resolution very close to the resolution obtained with one-photon excitation. These conclusions have been made on the basis of numerical results obtained by using a vector theory of diffraction. An experimental characterization of the resolution obtained with the 4Pi' microscope has to be made in order to confirm these predictions.

\end{document}